\documentclass[journal]{IEEEtran}
\usepackage{amsmath}
\usepackage{amssymb}
\usepackage{graphicx} 
\newtheorem{theorem}{Theorem}
\newtheorem{lemma}{Lemma}
\newtheorem{remark}{Remark}
\newtheorem{definition}{Definition}
\newtheorem{corollary}{Corollary}
\newtheorem{assumption}{Assumption}

\usepackage{color}

\IEEEoverridecommandlockouts 
\begin{document}
\title{Fault-Tolerant Control of  Linear Quantum Stochastic Systems}
\author{Shi Wang~~ Daoyi Dong\thanks{S.~Wang and D. Dong are with
School of Engineering and Information Technology, University of New South Wales, Canberra, Australia. peoplews3@hotmail.com (Shi Wang); daoyidong@gmail.com (Daoyi Dong). This work was  supported by the Australian Research Council's Discovery Projects funding scheme under Project DP130101658. }}
\maketitle
\begin{abstract}
In quantum engineering,  faults  may occur in
a quantum  control system, which will cause the quantum control system unstable or deteriorate other relevant performance of the system. This
note presents  an estimator-based fault-tolerant control design approach for a class of linear quantum stochastic systems subject to  fault signals. In this approach, the fault signals and some commutative components of the quantum system observables are estimated, and a fault-tolerant controller is designed to compensate the effect of the fault signals. Numerical procedures  are developed for controller design and an example is presented to demonstrate the proposed design approach.
\end{abstract}

\textbf{Index Terms---} Linear quantum stochastic system, quantum control, fault-tolerant control.
\section{Introduction}\label{sec:intro}
Developing quantum control theory has been  recognized a key task due to its potential application in emerging quantum technology \cite{WM10}, \cite{AT2012}, \cite{DP2010}. Some control methods such as optimal control \cite{DP2010}, $H^{\infty}$ control \cite{JNP08}, feedback control \cite{WM10}  have been  employed for enhancing the performance in quantum control systems. For practical quantum systems, the stochastic fluctuations in magnetic or electric fields or fault operations on the generators of quantum resources may introduce fault signals that will deteriorate the performance of quantum systems or result in instability \cite{DP2012}-\cite{GDI2015}. For example, classical (non-quantum) fault signals may originate from the change in environmental conditions (e.g., temperature) or the voltage fluctuation in controlling a laser generator.  It is thus expected to develop fault-tolerant control theory for quantum systems. Most classical fault-tolerant control methods \cite{A1976}-\cite{BKLS2006} cannot be straightforwardly applied to quantum control problems due to some unique characteristics of quantum systems such as measurement collapse and non-commutative observables \cite{DP2010}. However, some ideas of estimator-based design in classical control can be adapted to quantum control problems.  In  \cite{GDI2015},  fault-tolerant quantum filtering theory has been presented using the concept of quantum-classical conditional expectation. In this note, we present a systematic fault-tolerant control design method for a class of linear quantum stochastic systems subject to faults.
\begin{figure}
\centering
\includegraphics[width=80mm]{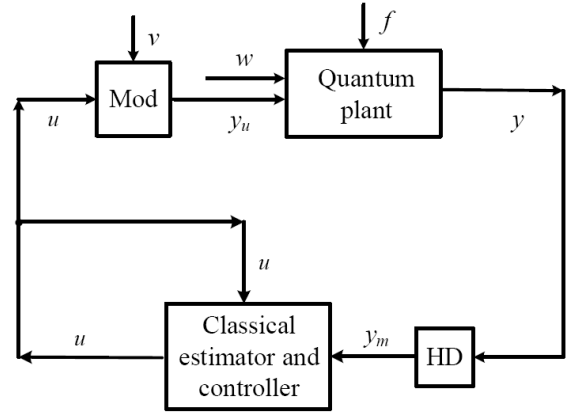}
\caption{A closed-loop  system with an estimator-based fault-tolerant controller. HD represents  a homodyne detector  for measurements and Mod
represents a modulator.   \label{fig:classical}}
\end{figure}
Linear quantum stochastic systems widely exist in quantum optics \cite{JNP08}, \cite{GZ04}, \cite{WM08}. Quantum optical components such as beam splitters, squeezers, phase shifters can be used to construct quantum networks for quantum information processing \cite{KLM01}. Some control methods involving coherent feedback \cite{JNP08}, \cite{NJP09} and measurement-based feedback \cite{WM10} have been used for enhancing the performance of linear quantum optical systems with uncertainties. The design problems of robust controllers or observers have also been investigated for uncertain linear quantum stochastic systems \cite{SPJ07}, \cite{Yama06}. For example, Yamamoto \cite{Yama06} presented the result of robust observer design for linear quantum systems. In particular, a class of linear quantum systems subject to time-varying norm-bounded parametric uncertainties was considered in \cite{Yama06} and a robust observer was proposed to guarantee the variance of the estimation error to be within a certain bound. Here, we develop an estimator-based approach for fault tolerant control design of linear quantum stochastic systems involving measurement-based feedback where we assume that the classical fault signal is independent of quantum noise. Different from uncertainties in the system Hamiltonian and the system operator considered in \cite{Yama06}, we consider the classical fault signals in quantum systems and design a controller to compensate the effect of fault signals. We aim to develop a fault-tolerant control design approach with a reduced-order dynamic estimator for a class of quantum systems subject to faults while Yamamoto \cite{Yama06} aimed at designing a robust full-order estimator for a class of quantum systems with uncertainties. The basic idea is illustrated in Fig. \ref{fig:classical},  where the meaning of different quantities will be explained in Section \ref{sec:closed}. A quantum optical plant subject to fault signals $f$  is measured using a homodyne detector (HD) \cite{WM10}. The output $y_m$ of the HD is used to establish a classical estimator and a classical controller for compensating the effect of $f$ on the quantum optical plant.


This note is organized as follows. Section \ref{sec:pre} introduces
some notations and gives a brief overview of linear quantum stochastic systems.
Section \ref{sec:closed} presents the setup of closed-loop
systems with input faults. The main results  for the  closed systems subject  to input faults are
provided in Section \ref{sec:criterion}. Section
\ref{Sec:classical-design} proposes numerical approaches for
the estimator-based fault-tolerant controller design  and  illustrates the proposed approaches using an example.
Section \ref{sec:conclusion} concludes this note.
\section{Preliminaries}
\label{sec:pre}
\subsection{Notation}
The notations used in this note are as follows:
$i=\sqrt{-1}$; the commutator
is defined by $[A, B]=AB-BA$. If $x$ and $y$ are column vectors of
operators, the commutator is defined by $[x, y^T]= xy^T -(yx^T)^T$.
If $X=[x_{jk}]$ is a matrix of linear operators or complex numbers,
then
$X^{\#} = [x^{*}_{jk}]$ denotes
the operation of taking the adjoint of each element of $X$, and
$X^{\dag}=[x^{*}_{jk}]^T$. $\parallel \!X\!\parallel$ represents Euclidean norm of $X$.  We
 define
$J={\small\left[\!\!\!
\begin{array}{cc}
0 & 1 \\
-1 & 0 \\
\end{array}
\!\!\!\right]}$, and
$\mathrm{diag}_{n}(M)$ denotes
a block diagonal matrix with a square matrix $M$ appearing $n$ times
on the diagonal block. The symbol $I_n$ denotes the $n \times n$
identity matrix. $0_{n\times m}$ denotes the $n\times m$ zero
matrix, where $n$ and $m$ can be determined from the context when the
subscript is omitted.

\subsection{Linear  quantum stochastic systems  and physical realizability} \label{QP}
Following \cite{JNP08}, an open quantum system  involving $n$ one degree of freedom open quantum
harmonic oscillators coupled to $m$ independent boson fields (e.g., optical beams) is described by linear differential equations of the form
\begin{align}\label{eq:fully-q}
dx(t)=&Ax(t)dt+Bdw(t)\nonumber,
\\
dy(t)=&Cx(t)dt+Ddw(t),
\end{align}
where $x = [q_1, p_1, q_2, p_2, \cdots, q_{n}, p_{n}]^T$ with position operators $q_j$ and momentum operators $p_j$
($j=1,\ldots,n$) describes a vector of self-adjoint possibly non-commutative system observables and the  non-commutation relation is defined as $xx^T-(xx^T)^T
= 2i \Theta$ with $\Theta = \mathrm{diag}_{n}(J)$. The boson fields $w=[w_{q_1}, w_{p_1}, w_{q_2}, w_{p_2}, \cdots, w_{q_{m}}, w_{p_{m}}]^T$ with analogous
field operators $ w_{q_k}(t)$, $w_{p_k}(t)$, ($k=1,\ldots,m$)  represent the input signal written as
\begin{equation} dw(t) dw(t) ^T=
F_{w}dt
\end{equation} with
$F_{w} = I_{2m} + i\mathrm{diag}_{m}(J)$. Commutation relations for the quantum field components of $w$
may be  denoted by:
\begin{equation*} [dw(t),
dw(t)^T] = (F_{w}- F_{w}^T)dt = 2i\Theta_{w}dt.
\end{equation*} Commutation relation  for the output $y$  is given by
\begin{equation*} [dy(t),
dy(t)^T] =2i\Theta_{y}dt.
\end{equation*}
The system matrices $A$, $B$, $C$ and
$D$ are real constant matrices of suitable dimension. The system matrices must satisfy  {\em physical realizability conditions} \cite{WNZJ2012}, \cite{WNZJ13} (i) $A\Theta+\Theta
A^{T}+B\Theta_{w}B^{T}=0$; (ii) $B\Theta_{w}D^T
 =-\Theta
C^T\label{condition-BCD}$; (iii) $D\Theta_{w}D^T=\Theta_{y}$.

\section{Model Description and Problem Formulation}\label{sec:closed}
This section presents a quantum optical plant with input faults, designs an estimator-based fault-tolerant controller  and introduces the
setup of a closed-loop  system.

Consider a quantum plant with fault signals described by
a non-commutative stochastic model  of the following form:
\begin{align}\label{original}
dx(t)=&Ax(t)dt+B_{w}dw(t)+B_{u}y_{u}(t)dt+B_ff(t)dt, \nonumber\\
dy(t)=&Cx(t)dt +Ddw (t),
\end{align}
where $A\in \mathbb{R}^{n\times n}$, $B_{w}\in
\mathbb{R}^{n\times n_{w}}$, $B_{u} \in \mathbb{R}^{n\times
n_{u}}$, $B_f \in \mathbb{R}^{n\times
n_{f}}$,  $C\in \mathbb{R}^{n_{y}\times n}$, $D\in
\mathbb{R}^{n_{y }\times n_{w }}$ ($n$, $n_{w}$, $n_u$ and $n_{y}$
 are even). $A$, $B=[B_{w} \ B_{u}]$, $C$ and $D$ should satisfy
{\em physical realizability conditions}. $x$ represents a vector of plant
variables and $w$ is the vector of vacuum quantum fields. $B_f$ is known and the real column vector $f(t)$ represents the unknown fault signal to be estimated. For example, $f(t)$ may originate from the voltage fluctuation in controlling the laser generator, malfunction of beam splitters, phase shifters, etc. The signal $y_u(t)$ is a control
input of the form
 \begin{align}\label{input-form}
 y_{u}(t)=&u(0)+\int^{t}_0u(s)ds+v(t),
\end{align}
where $v\in \mathbb{R}^{n_{u}}$ and $w$ are  independent and $n_{u}$ is even;  $u$ and $v$ are the signal and  quantum noise parts
of $y_u$,  respectively. When the quantum output signals $y(t)$ are
measured by homodyne detectors (HD), classical signals
$y_m(t)=Gy(t)\in \mathbb{R}^{n_{y_m}}$ are produced.  The matrix
$G$ corresponding to measurement processes satisfies the
condition below \cite{Nurdin2011}:
 \begin{align}\label{measurement-G}
G\Theta_{y}G^T=0
\end{align}
with $\mathrm{rank}(G)\leq \frac{n_{y}}{2}$, where  $G$   represents a static linear transformation (measurement processes) that converts boson fields into classical signals.

Now, we give the following
assumption and lemma in this work.

\begin{assumption}\label{assumption}
The fault signal $f(t)$ satisfies $\parallel f(t)\parallel \leq \alpha$ and $\parallel \dot{f}(t)\parallel \leq \beta$, where $\alpha>0$, $\beta>0$.
\end{assumption}

\begin{lemma}
Given a quantum optical plant with a fault signal  of the form \eqref{original}, there always exists a permutation matrix $T$ such that the transformed system is given as
\begin{align}\label{transformed}
d\tilde{x}(t)=&\tilde{A}\tilde{x}(t)dt+\tilde{B}_{w}dw(t)+\tilde{B}_{u}y_{u}(t)dt+\tilde{B}_ff(t)dt, \nonumber\\
dy(t)=&\tilde{C}\tilde{x}(t)dt +\tilde{D}dw (t)
\end{align}
with system matrices $\tilde{A}=TAT^{-1}=\left[
                                           \begin{array}{cc}
                                             \tilde{A}_{11}& \tilde{A}_{12}\\
                                             \tilde{A}_{21} &\tilde{A}_{22}\\
                                           \end{array}
                                         \right]
$, $\tilde{B}_{w}=TB_w=\left[
                                             \tilde{B}_{w_1}^T\
                                             \tilde{B}_{w_2}^T
                                         \right]^T$, $\tilde{B}_{u}=TB_u=\left[
                                             \tilde{B}_{u_1}^T  \
                                             \tilde{B}_{u_2}^T \right]^T$, $\tilde{B}_{f}=TB_f=\left[
                                             \tilde{B}_{f_1}^T  \
                                             \tilde{B}_{f_2}^T \right]^T$, $\tilde{C}=CT^{-1}=[\tilde{C}_1 \quad \tilde{C}_2]$, $\tilde{D}=D$ and new defined system variables $\tilde{x}(t)=Tx(t)=[\tilde{x}_{uo}(t)^T \quad\tilde{x}_{o}(t)^T]^T$, where $\tilde{x}_o\in \mathbb{R}^{n_o}$ represents $n_o$  components of $\tilde{x}(t)$ to be estimated while $\tilde{x}_{uo}\in \mathbb{R}^{n-n_o}$ represents unestimated components. Here $n_o\leq \frac{n}{2}$ and $\tilde{x}_o$ should satisfy $[\tilde{x}_o(t), \tilde{x}_o(t)^T]=0$ for all $t\geq 0$, so that the components of $\tilde{x}_o$ can be simultaneously observed.
\end{lemma}

Let $\eta=\left[\tilde{x}_{uo}^T\
                                         \tilde{x}_o^T \
                                         f^T
                                     \right]^T$ and $h(t)=\frac{df(t)}{dt}$. We first design an  augmented system for \eqref{transformed}  given by
\begin{align}\label{augmented-system}
d\eta(t)=&\mathcal{A}\eta(t)dt+\mathcal{B}_wdw(t)+\mathcal{B}_udy_u(t)+\mathcal{B}_hh(t)dt,\nonumber\\
dy_m(t)=&G(\mathcal{C}\eta(t)dt+\tilde{D}dw),
\end{align}
where the augmented matrices $\mathcal{A}=\left[
                                       \begin{array}{ccc}
                                        \tilde{A}_{11} &  \tilde{A}_{12}  & \tilde{B}_{f_{1} }\\
                                        \tilde{A}_{21}  &  \tilde{A}_{22}  & \tilde{B}_{f_{2} }\\
                                         0 & 0 & 0 \\
                                       \end{array}
                                     \right]$, $\mathcal{B}_w=\left[
                                       \begin{array}{c}
                                        \tilde{B}_{w_{1}} \\
                                        \tilde{B}_{w_{2}} \\
                                         0 \\
                                       \end{array}
                                     \right]$, $\mathcal{B}_u=\left[
                                       \begin{array}{c}
                                        \tilde{B}_{u_{1}}  \\
                                        \tilde{B}_{u_{2}}  \\
                                         0 \\
                                       \end{array}
                                     \right]$, $\mathcal{B}_h=\left[
                                         0  \
                                         0  \
                                        I
                                     \right]^T$, $\mathcal{C}=[\tilde{C}_1\    \tilde{C}_2\  0]$.

In order to  estimate plant observables $\tilde{x}_o$ and fault signal $f$ together,  we have from \eqref{augmented-system}
\begin{align}\label{reduced-system}
d\xi(t) =&\hat{A}\xi(t)dt+\hat{A}_{uo}\tilde{x}_{uo}(t)dt+\hat{B}_wdw(t)+\hat{B}_udy_u(t)+\nonumber\\
&\hat{B}_hh(t)dt,
\end{align}
where $\xi(t)=\left[\tilde{x}_o^T\
                                         f^T
                                     \right]^T\in
\mathbb{R}^{\hat{n}}$ with $\hat{n}=n_o+n_f$, $\hat{A}=\left[
                                                          \begin{array}{cc}
                                                            \tilde{A}_{22} & \tilde{B}_{f_{2} }\\
                                                            0 & 0 \\
                                                          \end{array}
                                                        \right]$, $\bar{A}_{uo}=\left[\begin{array}{c}
                                                            \tilde{A}_{21}\\
                                                            0 \end{array}
                                                        \right]\in
\mathbb{R}^{\hat{n}\times (n-n_o)}$, $\hat{B}_w=\left[
                                                           \tilde{B}_{w_{2}}^T \
                                         0 \right]^T\in
\mathbb{R}^{\hat{n}\times n_w}$, $\hat{B}_u=\left[\tilde{B}_{u_{2}}^T\
                                         0 \right]^T\in
\mathbb{R}^{\hat{n}\times n_u}$, $\hat{B}_h=\left[ 0 \
                                         I \right]^T\in
\mathbb{R}^{\hat{n}\times n_f}$.

We aim to build a classical linear estimator-based fault-tolerant controller for \eqref{reduced-system}  given by
\begin{align}\label{observer-system}
d\hat{\xi}(t)=&\hat{A}\hat{\xi}(t)dt+\hat{B}_uu(t)dt+L\left(dy_m(t)-G\hat{C}\hat{\xi}(t)dt\right),\nonumber\\
u(t)=&K\hat{\xi}(t),
\end{align}
where the estimate $\hat{\xi}(t)=[\hat{\tilde{x}}_o(t)^T \quad \hat{f}(t)^T]^T\in
\mathbb{R}^{\hat{n}}$, $\hat{C}=[\tilde{C}_2\quad 0]$ and $K=[K_x\quad K_f]$. Matrices $L$ and $K$ are gain parameters to be designed.

Define $e\!=\!\xi(t)\!-\!\hat{\xi}(t)\!=\![e_o(t)^T\quad e_f(t)^T]^T$ with $e_o(t)\!=\!\tilde{x}_o(t)\!-\!\hat{\tilde{x}}_o(t)$ and $e_f(t)=f(t)-\hat{f}(t)$. Now, we  have the error system:
\begin{align}\label{erro-equation}
\!de(t)\!=\!A_ee(t)dt+B_edw_e(t)\!+\!E\tilde{x}_{uo}(t)dt\!+\!\hat{B}_hh(t)dt,
\end{align}
where $A_e=\hat{A}- LG\hat{C}$, $B_e=[\hat{B}_w- LG\tilde{D}\quad \hat{B}_u]$, $E=\hat{A}_{uo}-LG\tilde{C}_1$ and $w_e(t)=\left[
                w(t)^T \ \
               v(t)^T \right]^T
$ satisfying $dw_e(t) dw_e(t) ^T=F_{w_e}dt$.

Interconnecting \eqref{transformed} and \eqref{erro-equation}, we obtain the following  closed system:
\begin{align}\label{interconnection-erro}
dz(t)=&\bar{A}z(t)dt+\bar{B}_wdw_e(t)+\bar{B}_ff(t)dt+\bar{B}_hh(t)dt,
\end{align}where \!$z(t)\!\!=\!\!\left[
              \begin{array}{c}
\tilde{x}(t)\\ e(t)\\
              \end{array}
            \right]$, $\bar{A}\!\!=\!\!\left[\!\!\! \begin{array}{ccc}
                   \tilde{A}_{11} \!& \! \tilde{A}_{12}\!+\!\tilde{B}_{u1}K_x&\!-\tilde{B}_{u1}K\!\\
                    \tilde{A}_{21}\! & \!\tilde{A}_{22}\!+\!\tilde{B}_{u2}K_x&-\tilde{B}_{u2}K\!\\
                    E & 0& A_e\\
                 \end{array}\!\!\! \right]=$\\
                 $\!\!\left[\!\!\!\!
                 \begin{array}{ccc}
                   \tilde{A}\!\!+\!\!\left[
                                 0 \ \ \tilde{B}_uK\left[
                                                     I \ \
                                                     0 \right]^T
                            \!\right]
                   &  \!-\tilde{B}_{u}K\!\\
                    \left[\!\begin{array}{cc}
                                 E & 0\\
                               \end{array}
                             \!\right] & A_e\!\!\\
                 \end{array}
               \!\!\!\right]$,  $\bar{B}_w\!\!=\!\!\!\left[\!\!
                 \begin{array}{cc}
                  \tilde{B}_{w1} &\tilde{B}_{u1}\\
                    \tilde{B}_{w2} &\tilde{B}_{u2}\\
                    \hat{B}_w\!\!-\!\!LG\tilde{D}&\tilde{B}_u\\
                 \end{array}
             \!  \!\right]\!\!=$\\$\!\!\left[\!\!
                 \begin{array}{cc}
                \!  \tilde{B}_{w} \!&\tilde{B}_{u}\!\\
                   \!\! \hat{B}_w\!\!-\!\!LG\tilde{D}&\!\tilde{B}_u\!\\
                 \end{array}
              \! \!\right]$, $\bar{B}_f\!\!=\!\!\left[\!
                                      \begin{array}{c}
                                        \!\!\tilde{B}_{f1}\!+\!\tilde{B}_{u1}K_f \!\! \\
                                        \!\!\tilde{B}_{f2}\!+\!\tilde{B}_{u2}K_f \!\! \\
                                        \!\!0\!\\
                                      \end{array}
                                    \!\right]\!=\!\left[\!\!
                                      \begin{array}{c}
                                        \!\!\tilde{B}_{f}\!+\!\tilde{B}_{u}K\left[\!\!
                                                                          \begin{array}{c}
                                                                           \! 0 \!\\
                                                                           \!\!I \!\\
                                                                          \end{array}
                                                                        \!\!\right]\!\!\!\\
                                                                                 0\!\!\\
                                      \end{array}
                                    \!\right]$, $\bar{B}_h=[0\quad 0\quad \hat{B}_h^T]^T$.  Here, $z(t)$ commutes with the fault signal $f(t)$.

\section{Main results}\label{sec:criterion}
In this section, we   present our main results (Theorem \ref{main theorem1} and Theorem \ref{Th:general-form}).  The following lemmas and definition   will be used in the proof of the main results.
\begin{lemma}\label{lemma1}
Given  arbitrary real column vectors $x$ and $y$ with the same dimension, the inequalities below hold
\begin{eqnarray}
xy^T+yx^T\leq xx^T+yy^T,\\
x^T y+y^Tx\leq x^Tx+y^Ty.
\end{eqnarray}
\end{lemma}
\begin{lemma}\label{lemma2}
Given an arbitrary
 real column vector $z$, if $\parallel \!z\!\parallel$$\leq \gamma$ ($\gamma>0$), then
\begin{eqnarray}
\gamma^2I-zz^T\geq 0.
\end{eqnarray}
\end{lemma}
\begin{lemma}\label{lemma-exponentially-stable}
If there  exists a real function  $g(t)$ of time $t$ satisfying the following relation
\begin{align}\label{lemma-eq1}
\frac{dg(t)}{dt}+cg(t)\leq \tau,
\end{align}
where $c$ and $\tau$ are positive real numbers, then inequality
\begin{align}\label{lemma-eq2}
g(t)\leq e^{-ct} g(0)+\frac{\tau}{c}
\end{align}
holds. That is, $g(t)$ is bounded for all $t>0$. When $\tau=0$, $\mathop{\mathrm{lim}}_{
t \rightarrow  \infty}g(t)=0$.
 \end{lemma}

The proof follows using a similar method in \cite{KR12}, \cite{JG10} and we omit the detailed proof.

 \begin{definition}\label{definition1}
 The system \eqref{interconnection-erro} is said to be {\em mean square bounded stable} if there exists a real  function  $r(t)=\langle V(t)\rangle$ satisfying inequality \eqref{lemma-eq2}, where $V(t)$ represents  an abstract internal energy
for the system \eqref{interconnection-erro} at time $t$.
 \end{definition}

 The following theorem relates the stability of system
\eqref{interconnection-erro} to certain linear matrix inequalities.
\begin{theorem}\label{main theorem1}
The system \eqref{interconnection-erro} is  bounded  stable in the sense of Definition \ref{definition1} with
\begin{align}\label{lypunov5}
\!\!\!\!\!\!\!\!g(t)\!\!=\!\!\langle V_{z}(t)\!\rangle
\!=\!\left\langle\!z(t)^TSz(t)\right\rangle
\end{align}
if there exists a real positive definite matrix
$S>0$ satisfying the following relation:
\begin{align}\label{u-condition}
\bar{A}^TS+ S\bar{A}+S\bar{B}_h\bar{B}_h^TS+S\bar{B}_f\bar{B}_f^TS\leq 0.
\end{align}
\end{theorem}
\quad \quad $\emph{Poof}$:
We construct a  Lyapunov function as $\langle V_z(t)\rangle=\left\langle\!z(t)^T Sz(t)\right\rangle$ with a real symmetric matrix $S$. Let $\Delta=\bar{A}^T S+ S\bar{A}+S\bar{B}_f\bar{B}_f^TS+S\bar{B}_h\bar{B}_h^TS$. Applying quantum It$\bar{o}$ rule to \eqref{lypunov5}, we have
\begin{align}\label{deriviation1}
&d\left\langle V_z(t)\right\rangle\nonumber\\
 =&\left\langle\!
dz(t)^TSz(t)+z(t)^TSdz(t)+dz(t)^TSdz(t)\right\rangle \nonumber\\
\!\!=&\left\langle z(t)^T\![\bar{A}^TS+ S\bar{A}+S\bar{B}_h\bar{B}_h^TS+S\bar{B}_f\bar{B}_f^TS]z(t)\right\rangle dt+\nonumber\\
&\left\langle h(t)^Th(t)+f(t)^Tf(t)+\mathrm{Tr}(\bar{B}_{w}^T S\bar{B}_{w}F_{w_e})\!\right\rangle dt-\nonumber\\
&\left\langle (h(t)-\bar{B}_h^TSz(t))^T(h(t)-\bar{B}_h^TSz(t))\right\rangle dt-\nonumber\\
&\left\langle (f(t)-\bar{B}_f^TSz(t))^T(f(t)-\bar{B}_f^TSz(t))\right\rangle dt\nonumber\\
\leq&\left\langle z(t)^T\![\bar{A}^TS+ S\bar{A}+S\bar{B}_h\bar{B}_h^TS+S\bar{B}_f\bar{B}_f^TS]z(t)\right\rangle dt+\nonumber\\
&\left\langle\alpha^2+\beta^2+\mathrm{Tr}(\bar{B}_{w}^T S\bar{B}_{w}F_{w_e})\!\right\rangle dt\nonumber\\
\leq&\lambda_{max}(\Delta)\left\langle z(t)^Tz(t)\right\rangle dt+\tau dt\nonumber\\
\leq&\frac{\lambda_{max}(\Delta)}{\lambda_{min}(S)}\left\langle z(t)^T Sz(t)\right\rangle dt+\tau dt
\nonumber\\
=&\frac{\lambda_{max}(\Delta)}{\lambda_{min}(S)}\langle {V_z(t)}\rangle dt+\tau dt,
\end{align}where $\tau=\alpha^2+\beta^2+\mathrm{Tr}(B_{w}^TSB_{w}F_{w_e})>0$; $\lambda_{min}$ and $\lambda_{max}$ represent the smallest eigenvalue of $S$ and the largest eigenvalue of $\Delta$, respectively. If \eqref{u-condition} holds and $S>0$, then we can conclude that $c=-\frac{\lambda_{max}(\Delta)}{\lambda_{min}(S)}>0$. From \eqref{deriviation1},
 the system \eqref{interconnection-erro} is bounded stable in the sense of Definition \ref{definition1}.
$\blacksquare$

\begin{theorem}\label{Th:general-form}
 Under Assumption \ref{assumption}, if there  exists a constant matrix  $P$, such that
\begin{align}\label{theorem1-error}
\left[
   \begin{array}{cc}
    \bar{A}P+P\bar{A}^T+4P & \bar{B} \\
    \bar{B}^T & -I \\
   \end{array}
 \right]
\leq0,
\end{align} with $\bar{B}\!=\![\sqrt{2}\alpha \bar{B}_f \  \sqrt{2}\beta \bar{B}_h \  \bar{B}_w]$, then \eqref{observer-system} can generate the estimate of $\xi$ satisfying
\begin{align}\label{them2-11}
\mathop{\mathrm{lim}}_{t\rightarrow\infty}\left\langle(\xi(t)-\hat{\xi}(t))(\xi(t)-\hat{\xi}(t))^T\right\rangle\leq \mathrm{Tr}(Y_2),
\end{align}
where $P=\left[
                                              \begin{array}{cc}
                                                Y_1 & N \\
                                                N^T &  Y_2\\
                                              \end{array}
                                            \right]$, $Y_{1}$ is a $n\times n$  symmetric matrix and $Y_2$ is a $\hat{n}\times \hat{n}$  symmetric matrix.
\end{theorem}
\quad \quad $\emph{Proof}$:
Define the symmetrized covariance matrix  $Q(t)=\frac{1}{2}\left\langle z(t)z(t)^T+\left(z(t)z(t)^T\right)^T\right\rangle$.
 By Lemmas \ref{lemma1} and \ref{lemma2}, and applying quantum  It$\bar{o}$  rules to $Q(t)$, we can obtain
\begin{align}\label{error1}
dQ(t)=&\frac{1}{2}\left\langle dz(t)z(t)^T+z(t)dz(t)^T+dz(t)dz(t)^T\right\rangle+\nonumber\\&\frac{1}{2}\left\langle \left(dz(t)z(t)^T+z(t)dz(t)^T+dz(t)dz(t)^T\right)^T\right\rangle\nonumber\\
=&\left\langle \bar AQ(t)+Q(t)\bar A^T+\frac{1}{2}B_w(F_v+F_v^T)B_w^T\right\rangle dt+\nonumber\\
&\frac{1}{2}\left\langle \bar{B}_f\left(f(t)z(t)^T+\left(z(t)f(t)^T\right)^T\right)\right\rangle dt+\nonumber\\
&\frac{1}{2}\left\langle\left(z(t)f(t)^T+\left(f(t)z(t)^T\right)^T\right)\bar{B}_f^T \right\rangle dt+\nonumber\\
&\frac{1}{2}\left\langle \bar{B}_h\left(h(t)z(t)^T+\left(z(t)h(t)^T\right)^T\right)+\right\rangle dt+\nonumber\\
&\frac{1}{2}\left\langle\left(z(t)h(t)^T+\left(h(t)z(t)^T\right)^T\right)\bar{B}_h^T \right\rangle dt\nonumber\\
\leq& \left\langle\bar AQ(t)+Q(t)\bar A^T+B_wB_w^T\right\rangle dt+\nonumber\\
&\left\langle z(t)z(t)^T+(z(t)z(t)^T)^T\right\rangle dt+\nonumber\\
&\left\langle\bar{B}_ff(t)(\bar{B}_ff(t))^T+(\bar{B}_ff(t)(\bar{B}_ff(t))^T)^T\right\rangle dt+\nonumber\\
&\left\langle z(t)z(t)^T+(z(t)z(t)^T)^T\right\rangle dt+\nonumber\\
&\left\langle\bar{B}_hh(t)(\bar{B}_hh(t))^T+(\bar{B}_hh(t)(\bar{B}_hh(t))^T)^T\right\rangle dt\nonumber\\
\leq& \left\langle\bar{A}Q(t)+Q(t)\bar{A}^T+4Q(t)+\bar B_w\bar B_w^T\right\rangle dt+\nonumber\\
&\left\langle \bar{B}_f(f(t)f(t)^T+(f(t)f(t)^T)^T)\bar{B}_f^T\right\rangle dt+\nonumber\\
&\left\langle\bar{B}_h(h(t)h(t)^T+(h(t)h(t)^T)^T)\bar{B}_h^T\right\rangle dt\nonumber\\
\leq& \left\langle(\bar{A}+2I)Q(t)+Q(t)(\bar{A}+2I)^T+\bar B_w\bar B_w^T\right\rangle dt+\nonumber\\
&\left\langle2\bar{B}_f\alpha^2I\bar{B}_f^T +2\bar{B}_h\beta^2I\bar{B}_h^T \right\rangle dt.
\end{align}
Hence, from \eqref{error1}, we have
\begin{align}\label{error2}
\dot{Q}(t)\leq&(\bar{A}+2I)Q(t)+Q(t)(\bar{A}+2I)^T+\bar B_w\bar B_w^T+\nonumber\\
&2\alpha^2\bar{B}_f\bar{B}_f^T +2\beta^2\bar{B}_h\bar{B}_h^T.
\end{align}

Let $\Gamma(t)=P-Q(t)$. If condition \eqref{theorem1-error} holds, we use Schur complements \cite{LMI} and it follows from \eqref{error2} that
\begin{align}\label{error-3}
\dot{\Gamma}(t)=&-\dot{Q}(t)\nonumber\\
\geq&-(\bar{A}+2I)Q(t)-Q(t)(\bar{A}+2I)^T-\bar B_w\bar B_w^T-\nonumber\\
&2\alpha^2\bar{B}_f\bar{B}_f^T-2\beta^2\bar{B}_h\bar{B}_h^T\nonumber\\
=&\!-\!(\bar{A}\!+\!2I)P\!-\!P(\bar{A}\!+\!2I)^T\!\!-\!\bar B_w\bar B_w^T\!-2\alpha^2\bar{B}_f\bar{B}_f^T \!-\nonumber\\&2\beta^2\bar{B}_h\bar{B}_h^T\!+
(\!\bar{A}\!+\!2I\!)\Gamma(t)\!+\!\Gamma(t)\!(\!\bar{A}\!+\!2I\!)^T\nonumber\\
\geq&(\bar{A}+2I)\Gamma(t)+\Gamma(t)(\bar{A}+2I)^T.
\end{align}
From \eqref{error-3}, we obtain $\Gamma(t)\geq e^{(\bar{A}+2I)t}\Gamma(0)e^{(\bar{A}+2I)^Tt}$. If condition \eqref{theorem1-error} holds, the matrix $\bar{A}+2I$ is Hurwitz. It can be thus seen that $\mathop{\mathrm{lim}}_{t\rightarrow\infty}\Gamma(t)\geq 0$. Then, we have $\mathop{\mathrm{lim}}_{t\rightarrow\infty}Q(t)\leq P$, which implies \eqref{them2-11}.
\quad\quad\quad\quad\quad\quad\quad\quad\quad\quad\quad\quad\quad \quad \quad \quad  $\blacksquare$

From the  conclusions of Theorem \ref{main theorem1} and Theorem \ref{Th:general-form},  we have the following corollary.
\begin{corollary}\label{corollary1}
If the following relation holds
\begin{align}\label{corollary1-condition}
4P\!+\!(2\alpha^2\!-\!1)\bar{B}_f\bar{B}_f^T\!+\!(2\beta^2\!-\!1)\bar{B}_h\bar{B}_h^T\!+\!\bar{B}_w\bar{B}_w^T\geq \!0,
 \end{align}\label{error3} then condition \eqref{theorem1-error} implies \eqref{u-condition} for $S=P^{-1}$.
\end{corollary}

\quad \quad $\emph{Proof}$:
Let
$$W_1=P\bar{A}^T+\bar{A}P+\bar{B}_h\bar{B}_h^T+\bar{B}_f\bar{B}_f^T$$ and
$$W_2=\bar{A}P+P\bar{A}^T+4P+\bar{B}\bar B^T.$$
Note that $\bar{B}\!=\![\sqrt{2}\alpha \bar{B}_f \  \sqrt{2}\beta \bar{B}_h \  \bar{B}_w]$. We have
$$\bar{B}\bar B^T=2\alpha^2\bar{B}_f\bar{B}_f^T+2\beta^2\bar{B}_h\bar{B}_h^T+\bar{B}_w\bar{B}_w^T.$$
When the relation \eqref{corollary1-condition} holds, we have $$W_1\leq W_2.$$
If the condition \eqref{theorem1-error} is satisfied, then $$W_2\leq 0.$$
Hence, $$W_1\leq 0.$$
For $S>0$, it is clear that $$SW_1S\leq 0.$$
When $S=P^{-1}$, from $SW_1S\leq 0$ we have
$$\bar{A}^TS+ S\bar{A}+S\bar{B}_h\bar{B}_h^TS+S\bar{B}_f\bar{B}_f^TS\leq0. $$
That is, the relation \eqref{u-condition} holds. \quad \quad \quad \quad \quad \quad \quad \quad \quad \quad \quad $\blacksquare$

In the next section, we will focus on \textbf{Problem} \textbf{1} and provide numerical procedures to solve the problem.

\textbf{Problem} \textbf{1}: \ Given a quantum optical plant with faults of the form \eqref{original} that can be transformed into \eqref{transformed} and  for an estimation
error upper bound $\gamma$ (expected to be close to $0$, i.e., small error bound), find a classical linear estimator-based fault-tolerant controller of the form \eqref{observer-system}   with parameters $L$ and $K$ such that the following conditions  hold for fixed $G$ satisfying \eqref{measurement-G}:
 \begin{enumerate}

\item There exists a symmetric matrix  $P>0$ satisfying \eqref{theorem1-error} and \eqref{corollary1-condition}.

\item $0<\mathrm{Tr}(Y_2)\leq\gamma$.

\end{enumerate}

\section{Estimator-based fault-tolerant controller synthesis}\label{Sec:classical-design}
In this section, we  propose   numerical procedures  for  controller design to solve \textbf{Problem} \textbf{1} and then present an example to illustrate the proposed method.

Note that the designed parameters $K$ and $L$ are embedded in the system matrices of \eqref{interconnection-erro}.   To design the two parameters, we  extend the method proposed in \cite{SGCr97}, \cite{NJP09} by introducing auxiliary variables $N\in
\mathbb{R}^{n\times\hat{n}}$, $M_1\in
\mathbb{R}^{n\times \hat{n}}$, $M_2\in
\mathbb{R}^{\hat{n}\times n}$, $X_1\in
\mathbb{R}^{n\times n}$, $Y_1\in
\mathbb{R}^{n\times n}$, $Y_2\in
\mathbb{R}^{\hat{n}\times \hat{n}}$, where
$NM_2+Y_1X_1=I_n$, $N^TX_1+Y_2M_2=0$;
$X_1$, $Y_1$ and $Y_2$ are symmetric. Assuming that $P=\left[
                                              \begin{array}{cc}
                                                Y_1 & N \\
                                                N^T &  Y_2\\
                                              \end{array}
                                            \right]$ and $\Pi=\left[
                                                                 \begin{array}{cc}
                                                                   X_1 &M_1 \\
                                                                   M_2 & 0 \\
                                                                 \end{array}
                                                               \right]$, we have $P\Pi=\left[
                                              \begin{array}{cc}
                                                I_n & Y_1M_1 \\
                                                 0& N^TM_1\\
                                              \end{array}
                                            \right]$.

Performing congruence transformations on  inequality
\eqref{theorem1-error} with transformation matrix $\mathrm{diag}(\Pi,
I)$, we have
\begin{align}
\!\!\!&\left[
   \begin{array}{cc}
   \Pi^T & 0 \\
    0 & I \\
   \end{array}
 \right] \left[
   \begin{array}{cc}
    \bar{A}P+P\bar{A}^T+4P & \bar{B} \\
    \bar{B}^T & -I \\
   \end{array}
 \right]\left[
   \begin{array}{cc}
   \Pi &  0  \\
    0  & I  \\
   \end{array}
 \right]\nonumber\\
   =&\left[
   \begin{array}{cc}
   \Pi^T\left(\bar{A}P+P\bar{A}^T+4P\right)\Pi & \Pi^T\bar{B} \\
    \bar{B}^T\Pi & -I \\
   \end{array}
 \right]\nonumber\\
\!\!\!\!\!\!\!\!&\!\!\!\!=\!\!\left[\!\begin{array}{c}
     \! \!\! \Omega_1\!+\!\Omega_1^T+4X_1\quad \quad \  \ \ \mathbf{A}\!+\!\Omega_2^T\!+\!4M_1\quad \quad   \Omega_4\!\!\\
\!\!\!\mathbf{A}^T\!+\!\Omega_2\!+\!4M_1^T\quad  \Omega_3\!+\!\Omega_3^T\!+\!4M_1^TY_1M_1\quad  \Omega_5\!\!\\
\quad\quad \quad    \Omega_4^T  \quad\quad\quad\quad\quad\quad \Omega_5^T \quad \quad\quad\quad\quad   -I\!\!\\
      \end{array}
   \!\! \right]\!\!<0,
 \label{LMI-condition-classical1}
\end{align}
        where $\bar{B}\!=\![\sqrt{2}\alpha \bar{B}_f \ \ \sqrt{2}\beta \bar{B}_h \ \ \bar{B}_w]$, $\Omega_1=X_1\tilde{A}+\left[0 \  X_1\tilde{B}_uK_x\right]\!+\!M_2^T\left[
                                              E \  0 \right]$, $\Omega_2\!=\!M_1^T\tilde{A}\!+\!\left[0 \ M_1^T\tilde{B}_uK_x \right]$, $\Omega_3\!=\!M_1^T\tilde{A}Y_1M_1\!+\!\left[0 \  M_1^T\tilde{B}_uK_x \right]Y_1M_1-M_1^T\tilde{B}_uKN^TM_1$, $\Omega_4\!=\!\sqrt{2}\alpha(X_1\tilde{B}_f\!+\!X_1\tilde{B}_uK_f)\!+\!\sqrt{2}\beta M_2^T\hat{B}_h\!+\!X_1[\tilde{B}_{w}\ \tilde{B}_u]\!+\!M_2^T[\hat{B}_{w}-LG\tilde{D}\  \tilde{B}_u]$,
$\Omega_5\!=\!\sqrt{2}\alpha(M_1^T\tilde{B}_f\!+\!M_1^T\tilde{B}_uK_f)\!+\!M_1^T[\tilde{B}_{w}\  \tilde{B}_u]$,
                                          $\mathbf{A}\!=\!X_1\tilde{A}Y_1M_1\!+\!\left[
                                              0 \  X_1\tilde{B}_uK_x \right]Y_1M_1\!+\!M_2^T\left[E \  0 \right]Y_1M_1-X_1\tilde{B}_uKN^TM_1-M_2^TA_eN^TM_1$.
\subsection{Numerical procedure for controller  design}\label{Controller-design}
$\mathbf{Case}$ $\mathbf{1}$ $n\geq \hat{n}$

To replace nonlinear entries in
\eqref{LMI-condition-classical1} by linear ones, we need to introduce appropriate matrix lifting
variables and the associated equality constraints. Let $Z_{x_1}\!=\!X_1, Z_{x_2}\!=\!Y_1, Z_{x_3}\!=\!\mathbf{M}_1^T\!=\!\left[M_1\
                                         0
                                     \right]^T
, Z_{x_4}\!=\!\mathbf{M}_2^T\!$$=\!\left[M_2^T\  0\right]$, $Z_{x_5}=\mathbf{N}^T=\left[
                                                                      N\
                                                                      0
                                                                  \right]^T
$,
$Z_{x_{6}}=\mathbf{L}=\mathrm{diag}\{L^T, 0\}
, Z_{x_7}=\mathbf{K}^T=\mathrm{diag}\{K^T, 0\}, Z_{x_8}=\mathbf{Y}_2=\mathrm{diag}\{Y_2, 0\}
, Z_{x_9}=\mathbf{N}=[N\  0]$. Define a symmetric matrix $Z$
of dimension $29n\times 29n$ as $Z\!=\!\mathbf{V}\!\mathbf{V}^T$, where
$\mathbf{V}=[I_n\  Z_{x_1}^T \cdots\  Z_{x_{9}}^T\
Z_{v_1}^T \  Z_{v_2}^T  \cdots\  Z_{v_{20}}^T]^T$, $Z_{v_1}\!=\!\mathbf{M}_1^T\left[\tilde{B}_u\  0\right]$, $Z_{v_2}\!=\!\mathbf{M}_1^T\left[\tilde{B}_u\  0\right]\mathbf{K}$, $Z_{v_3}\!=\!\mathbf{M}_1^T\tilde{A}\!+\!\left[\!0 \  \mathbf{M}_1^T\![\tilde{B}_u\ 0]\mathbf{K}[I_{n_{o}}\  0]^T\!\right]$,\\ $Z_{v_4}\!=\!\mathbf{M}_{1}^T\!Y_1^T, Z_{v_{5}}\!=\!\left(\!\mathbf{M}_1\!\tilde{A}\!+\!\!\left[0 \  \mathbf{M}_1^T\!\left[\tilde{B}_u\  0\right]\!\mathbf{K}\!\left[\!\begin{array}{c}\!\! I_{n_{o}}\! \\
                                                                                                                         \!\!0\! \\
                                                                                                                       \end{array}
                                                                                                                     \!\right]\!\right]\!\right)\!Y_1\mathbf{M}_{1}$,
$Z_{v_{6}}\!=\!\mathbf{M}_1^T\!\!\left[\tilde{B}_u\ 0\right]\mathbf{K}\mathrm{diag}\{ I_{\hat{n}}, 0\}, Z_{v_{7}}=\mathbf{M}_1^T\!\mathbf{N}, Z_{v_{8}}\!=\!\mathbf{M}_1^T\times$\\$\left[\!\tilde{B}_u\ 0\!\right]\!\mathbf{K}\mathrm{diag}\{I_{\hat{n}}, 0\}\!\mathbf{N}^T\!\mathbf{M}_1, Z_{v_{9}}\!\!\!=\!\!X_1\!\!\!\left[\!\tilde{B}_u\  0\right]\!, Z_{v_{10}}\!\!=\!\!X_1\!\!\!\left[\!\tilde{B}_u\ 0\!\right]\!\!\mathbf{K}$,\\ $Z_{v_{11}}\!=\!\mathbf{M}_2^T\!\mathrm{diag}\!\{ I_{\hat{n}}, 0\}$, $Z_{v_{12}}\!=\!\!\mathbf{M}_2^T\!\mathrm{diag}\{I_{\hat{n}}, 0\}\mathbf{L}, Z_{v_{13}}\!\!=\!\!X_1\!\tilde{A}+\\\!\left[\!0\   X_1[\!\tilde{B}_u\  0]\mathbf{K}\!\left[\!\!\!\begin{array}{c}
                                                                                                                                                                                                               I_{\hat{n}} \\
                                                                                                                                                                                                               0\\
                                                                                                                                                                                                             \end{array}
                                                                                                                                                                                                           \!\!\!\right]\!\right]\!+\!\left[\!\left(\!\mathbf{M}_2^T\!\left[\!\!\!
                                                                                                                                                                                                             \begin{array}{c}
                                                                                                                                                                                                               I_{\hat{n}} \\
                                                                                                                                                                                                               0\\
                                                                                                                                                                                                             \end{array}
                                                                                                                                                                                                           \!\!\!\right]
                                                                                                                                                    \!\!\!\hat{A}_{uo}\!\!-\!\!\mathbf{M}_2^T\!\left[\!\!
                                                                                                                                                                                                 \begin{array}{c}
                                                                                                                                                                                                   I_{\hat{n}} \ 0 \\
                                                                                                                                                                                                  0 \ \ 0 \\
                                                                                                                                                                                                 \end{array}
                                                                                                                                                                                               \!\!\right]
                                                                                                                                                    \mathbf{L}G\tilde{C}_1\!\right) \ 0\!\right]$, $Z_{v_{14}}\!=\!X_1\tilde{A}Y_1\mathbf{M}_{1}+\!\!\left[\!0 \ X_1[\tilde{B}_u\ \ 0]\mathbf{K}\mathrm{diag}\{ I_{n_{o}}, 0\}\right]Y_1\mathbf{M}_{1}\!\!+\!\! \left[\!\left(\!\mathbf{M}_2^T\!\left[I_{\hat{n}} \ 0 \right]^T \hat{A}_{uo}\!\!-\!\!\mathbf{M}_2^T\!\mathrm{diag}\{ I_{\hat{n}}, 0\}\mathbf{L}G\tilde{C}_1\!\right) \ 0\right]\!Y_1\mathbf{M}_{1}$, $Z_{v_{15}}\!=\!X_1[\tilde{B}_u\ 0]\mathbf{K}+\mathbf{M}_2^T\mathrm{diag}\{ \hat{A}, 0\}-\mathbf{M}_2^T \mathrm{diag}\{ I_{\hat{n}}, 0\}\mathbf{L}G\hat{C}[I_{\hat{n}}\quad 0]$, $Z_{v_{16}}\!=\!(X_1[\tilde{B}_u\ 0]\mathbf{K}\mathbf{N}^T\mathbf{M}_1+\mathbf{M}_2^T\mathrm{diag}\{ \hat{A}, 0\}\mathbf{N}^T\mathbf{M}_1-\mathbf{M}_2^T \mathrm{diag}\{ I_{\hat{n}}, 0\}\mathbf{L}G\hat{C}\!\left[I_{\hat{n}}\ 0\right]\!\mathbf{N}^T\!\mathbf{M}_1$, $Z_{v_{17}}\!\!=\!\!X_1\!\mathbf{N}$, $Z_{v_{18}}\!\!=\!\!\mathbf{M}_2^TY_2$, \\ $Z_{v_{19}}\!=\!\mathbf{N}\mathbf{M}_2$, $Z_{v_{20}}\!=\!Y_1X_1$, $Z_{v_{21}}=\mathbf{M}_1^TY_1\mathbf{M}_1$.

The symmetric matrix
$Z$ should satisfy the following conditions:
\begin{align}\label{Z-conditions}
&\!Z\!\!\geq\!0;  Z_{0, 0}\!\!-\!\!I_{n\times n}\!\!=\!0;
Z_{v_1}\!\!\!-\!\!Z_{x_3}\tilde{B}_u\left[I_{n_o}\  0\right]\!\!=\!0;
Z_{v_2}\!\!\!-\!\!Z_{v_1}\!Z_{x_7}^T\!\!=\!0;\nonumber\\
&\!\!Z_{v_3}\!\!\!-\!\!Z_{x_3}\tilde{A}\!\!-\!\!\!\left[0 \ Z_{v_2}\!\left[\!
                                                             \begin{array}{c}
                                                              \!\!\! I_{n_{o}}\!\!\!\\
                                                              \!\!\!0\!\!\! \\
                                                             \end{array}\!\right]
\!\right]\!\!=\!0;
Z_{v_4}\!\!-\!\!Z_{x_3}Z_{x_2}^T\!\!=\!0; Z_{v_5}\!\!\!-\!\!Z_{v_3}Z_{v_4}^T\!\!=\!0;\nonumber\\
&\!Z_{v_6}\!\!-\!Z_{v_2}\mathrm{diag}\{ I_{\hat{n}}, 0\}\!=\!0;  Z_{v_7}\!-\!Z_{x_3}Z_{x_5}^T\!=\!0;  Z_{v_8}\!-\!Z_{v_6}Z_{v_7}^T\!=\!0;\nonumber\\
&\!\!Z_{v_{9}}\!\!\!-\!\!Z_{x_1}\![\tilde{B}_u\  0]\!=\!0;  Z_{v_{10}}\!\!\!-\!\!Z_{v_9}Z_{x_7}^T\!=\!0; Z_{v_{11}}\!\!\!-\!\!Z_{x_{4}}\mathrm{diag}\{ I_{\hat{n}}, 0\}\!\!=\!0;\nonumber\\
&\!\!Z_{v_{12}}\!\!-\!\!Z_{v_{11}}Z_{x_6}^T\!\!=\!0;
Z_{v_{13}}\!\!-\!\!\left[\!Z_{x_4}\left[I_{\hat{n}} \ 0 \right]^T\!\!\!\hat{A}_{uo}\!\!-\!\!Z_{v_{12}}\!\left[\!I_{\hat{n}} \ 0 \right]^T\!\!G\tilde{C}_1\  0\right]\!\!-\nonumber\\
&\! Z_{x_1}\tilde{A}\!-\!\!\left[0\  Z_{v_{10}}\!\left[I_{n_o} \ 0 \right]^T \!\right]\!\!=\!0;  Z_{v_{14}}\!-\!Z_{v_{13}}Z_{v_4}^T\!\!=\!0; Z_{v_{15}}\!-\!Z_{v_{10}}-\nonumber\\
&\!\!Z_{x_4}\mathrm{diag}\{\hat{A}, 0\}\!+\!Z_{v_{12}}G\hat{C}[I_{\hat{n}}\ 0]\!=\!0; Z_{v_{16}}\!-\!Z_{v_{15}}Z_{v_{7}}^T\!=\!0; Z_{v_{17}}\!-\nonumber\\
&\!\!Z_{x_4}Z_{x_5}^T\!=\!0; Z_{v_{18}}\!\!-\!\!Z_{x_4}Z_{x_8}^T\!=\!0; Z_{v_{19}}\!-\!Z_{x_5}\!Z_{x_4}^T\!=\!0; Z_{v_{20}}\!\!-\!\!Z_{x_2}Z_{x_1}^T\!\nonumber\\
&\!\!=0; Z_{v_{21}}\!-\!Z_{x_3}Z_{v_4}^T\!=0; Z_{v_{19}}\!+\!Z_{v_{20}}\!=\!I_n;Z_{x_1}\!-\!Z_{x_1}^T\!=\!0; Z_{x_2}\!-\nonumber \\
&\!\!Z_{x_2}^T\!=\!0;  Z_{x_8}\!-\!Z_{x_8}^T\!=\!0; Z_{x_5}\!-\!Z_{x_9}^T\!=\!0;  [0\quad I_{n\!-\!\hat{n}}]Z_{x_3}\!=\!0; Z_{x_4}\times\nonumber\\
&\!\![0\ I_{n\!-\!\hat{n}}]^T\!=\!0; [I_{n_{y_m}}\  0]Z_{x_6}[0\  I_{n-\hat{n}}]^T\!=\!0; [0\  I_{n-n_{y_m}}]Z_{x_6}\!=\!0; \nonumber\\
\!&\!\![I_{\hat{n}}\ 0]Z_{x_7}[0\ I_{n-n_u}]^T=0; \  [0\  I_{n\!-\!\hat{n}}]Z_{x_7}\!=\!0; [0\  I_{n-\hat{n}}]Z_{x_8}\!=\!0;\nonumber\\
\!&\!\![I_{\hat{n}}\  0]\!Z_{x_8}\!\!\left[0\  I_{n\!-\!\hat{n}}\right]^T\!\!=\!0; Z_{x_9}\![0\  I_{n\!-\!\hat{n}}]^T\!\!=\!\!0; Z_{v_{17}}\!\!+\!\!Z_{v_{18}}\!\!=\!0;
\end{align} and a rank constraint
\begin{align}\label{classical-n_oank-conditions}
\mathrm{rank}(Z)\leq n.
\end{align}
Condition \eqref{corollary1-condition} is satisfied with
\begin{align}
&P=\left[
                                              \begin{array}{cc}
                                                Z_{x_2} & ([I_{\hat{n}}\quad 0]Z_{x_5})^T \\
                                               \ [I_{\hat{n}}\quad 0]Z_{x_5} & \quad [I_{\hat{n}}\quad 0]Z_{x_8}[I_{\hat{n}}\quad 0]^T\\
                                              \end{array}
                                            \right]>0
\end{align} and \eqref{LMI-condition-classical1} is satisfied with $\Omega_1=Z_{v_{13}}, \Omega_2=\left[I_{\hat{n}}\ \ 0\right]Z_{v_{3}}, \Omega_3=\\\left[I_{\hat{n}}\  0\right]\!\!\left(Z_{v_{5}}\!\!-\!\!Z_{v_{8}}\right)\!\left[I_{\hat{n}} \ 0 \right]^T\!\!, $$\ \Omega_4\!\!=\!\!\sqrt{2}\alpha\!\!\left(\!Z_{x_1}\!\tilde{B}_f\!+\!\!Z_{v_{10}}\!\left[0\  I_{n_f} \ 0 \right]^T\right)$$\!+\!\!\!\\\left[Z_{x_4}\!\!\left[I_{\hat{n}} \ 0 \right]^T\!\!\hat{B}_{w}\!-\!Z_{v_{12}}\!\left[I_{n_{y_m}} \ 0 \right]^T\!\!\!G\!\tilde{D}\ \ \ Z_{x_4}\!\left[I_{\hat{n}} \
                                                                                                                  0 \right]^T\!\!\tilde{B}_u\!\right]\!+ \sqrt{2}\beta\!\times\\Z_{x_4} \left[
                                                                                                                  I_{\hat{n}} \ 0 \right]^T\hat{B}_h\!+\!Z_{x_1}[\tilde{B}_{w}\ \tilde{B}_u]$, $\Omega_5\!=\!\sqrt{2}\alpha[I_{\hat{n}}\  0]Z_{x_3}\tilde{B}_f\!+\!\sqrt{2}\alpha Z_{v_2}\left[0\  I_{n_f} \ 0 \right]^T+[I_{\hat{n}}\ 0]Z_{x_3}[\tilde{B}_w\ \tilde{B}_u]$, $\mathbf{A}=(Z_{v_{14}}+Z_{v_{16}})\left[
                                                                                                                  I_{\hat{n}} \   0 \right]$. We expect to find an estimation
error bound $\gamma$  close to $0$ to satisfy the following condition:
\begin{align}\label{bound}
0<\mathrm{Tr}\left(\left[I_{\hat{n}} \ \  0 \right]Z_{x_8}\left[I_{\hat{n}} \ \  0 \right]^T\right) \leq\gamma.
\end{align}

If we can employ  semi-definite programming to solve the
feasibility problem with constraints \eqref{corollary1-condition}-\eqref{bound}
in which decision variables are the elements of $\mathbf{V}$ (see  \cite{Orsi06}, \cite{LJ04}, \cite{Orsi05}),
then we have
\begin{align*}
&L=[I_{\hat{n}}\  0]Z_{x_{6}}^T\left[ I_{n_{y_m}} \ 0\right]^T,\\
&K=[I_{n_u}\  0]Z_{x_{7}}^T\left[I_{\hat{n}} \  0  \right]^T.
\end{align*}

$\mathbf{Case}$ $\mathbf{2}$ $n<\hat{n}$

Similarly, we let $Z_{x_1}=\mathbf{X}_1=\mathrm{diag}\{X_1, 0\}$,
$Z_{x_2}=\mathbf{Y}_1=$\\$\mathrm{diag}\{Y_1, 0\}$, $Z_{x_3}\!=\!\mathbf{M}_1^T\!=\![M_1^T\  0]$, $Z_{x_4}\!=\!\mathbf{M}_2^T\!=\!\left[
                                         M_2\
                                         0 \right]^T$, $Z_{x_5}\!=\!\mathbf{N}^T\!\!=\!\left[N^T\  0\right]
$,
$Z_{x_{6}}\!=\!\mathbf{L}^T\!=\!\!\left[L \
               0 \right]^T\!, Z_{x_7}\!=\!\mathbf{K}^T\!=\![K^T\ 0]$,\\ $Z_{x_8}=Y_2$, $Z_{x_9}=\mathbf{N}=\left[
              N^T \
              0\right]^T$. Then we define a symmetric matrix $Z$
of dimension $27\hat{n} \times 27\hat{n}$ as $Z= \mathbf{V}\mathbf{V}^T$, where
$\mathbf{V}=[I_n\  Z_{x_1}^T\ \cdots\  Z_{x_{9}}^T$  $\
Z_{v_1}^T  \ \cdots\  Z_{v_{18}}^T]^T$, $Z_{v_1}\!=\!\mathbf{M}_1^T\mathrm{diag}\{\tilde{B}_{u}, 0\}$,
 $Z_{v_2}\!=\!\mathbf{M}_1^T\mathrm{diag}\{\tilde{B}_{u}, 0\}\mathbf{K}$, $Z_{v_3}\!=\!\mathbf{M}_{1}^T\mathbf{Y}_1^T$, $Z_{v_4}=[\mathbf{M}_{1}^T\!\tilde{A}\ \ 0]+\left[\left[0 \  \mathbf{M}_1^T\mathrm{diag}\{\tilde{B}_{u}, 0\}\mathbf{K}\left[
                                         I_{n_{o}}\
                                         0 \right]^T\right]
                                                                                        \  0\right]$, $Z_{v_{5}}=\mathbf{M}_{1}^T\mathbf{Y}_1^T$, $Z_{v_6}=\left([\mathbf{M}_{1}^T\!\tilde{A}\ 0]+\left[\left[0 \  \mathbf{M}_1^T\mathrm{diag}\{\tilde{B}_{u}, 0\}\mathbf{K}\left[
                                         I_{n_{o}}\
                                         0 \right]^T\right]\  0\right]\right)
\mathbf{Y}_1\mathbf{M}_1$,
$\!Z_{v_{6}}\!\!=\!\!\mathbf{M}_1^T\mathbf{N}$, $Z_{v_{7}}\!=\!\mathbf{M}_1^T\!\mathrm{diag}\{\tilde{B}_{u}, 0\}
           \mathbf{N}^T\!\mathbf{M}_1$,
$Z_{v_{8}}\!=\!\mathbf{X}_1\left[\!
                         \begin{array}{c}
                           \!\!\!\tilde{B}_{u} \ 0\!\!\! \\
                          \!\!\! 0 \ \ \ 0 \!\!\!\\
                         \end{array}
                       \right]$,\\ $Z_{v_{9}}\!=\!\mathbf{X}_1\mathrm{diag}\{\tilde{B}_{u}, 0\}\mathbf{K}$, \!$Z_{v_{10}}\!=\!\mathbf{M}_2^T\mathbf{L}$, $Z_{v_{11}}\!\!=\!\mathbf{X}_1\!\mathrm{diag}\{\tilde{A}, 0\}\!+$\\$\!\left[0\  \mathbf{X}_1\!\mathrm{diag}\{\tilde{B}_{u}, 0\}\mathbf{K}[0\  I_{n_f}]^T\ 0\right]\!\!+\!\!\left[\!\left[\!\mathbf{M}_2^T\!\hat{A}_{uo}\!\!-\!\!\mathbf{M}_2^T\!\mathbf{L}\!G\tilde{C}_1\  0 \!\right]\  0\right]$,\\
                                                                                      $Z_{v_{12}}=\bigg(\!\mathbf{X}_1\mathrm{diag}\{\tilde{A}, 0\}\!+\!\left[0\ \mathbf{X}_1\mathrm{diag}\{\tilde{B}_{u}, 0\}\mathbf{K}[0\ I_{n_f}]^T\ 0\right]\!\!+\!\left[\!\left[\!\mathbf{M}_2^T\!\hat{A}_{uo}\!\!-\!\!\mathbf{M}_2^T\!\mathbf{L}G\tilde{C}_1\ 0 \!\right]\  0\!\right]\!\!\!\bigg)\!\mathbf{Y}_1\!\mathbf{M}_1$,
                                                                                        $Z_{v_{13}}\!\!=\!\mathbf{X}_1\mathrm{diag}\{\tilde{B}_{u}, 0\}\mathbf{K}\!+$\\$\mathbf{M}_2^T\hat{A}-\mathbf{M}_2^T\mathbf{L}G\tilde{C}$,
                                                                                        $Z_{v_{14}}=(\mathbf{X}_1\mathrm{diag}\{\tilde{B}_{u}, 0\}\mathbf{K}+\mathbf{M}_2^T\hat{A}-\mathbf{M}_2^T\mathbf{L}G\tilde{C})\mathbf{N}^T\mathbf{M}_1$, $Z_{v_{15}}\!=\!\mathbf{X}_1\mathbf{N}$, $Z_{v_{16}}\!=\!\mathbf{M}_2Y_2$, $Z_{v_{17}}\!=\!\mathbf{N}\mathbf{M}_2$, $Z_{v_{18}}\!=\!\mathbf{Y}_1\mathbf{X}_1$, $Z_{v_{18}}=\mathbf{M}_1^T\mathbf{Y}_1\mathbf{M}_1$.

The symmetric matrix
$Z$ should satisfy the following conditions:
\begin{align}\label{Z-2-bound}
&Z\geq 0;\  Z_{0, 0}-I_{\hat{n}\times \hat{n}}=0; \ Z_{v_1}\!-Z_{x_3}\!=\!0; \ Z_{v_2}\!-Z_{v_1}Z_{x_7}^T=0;\nonumber \\
&\mathrm{diag}\{\tilde{B}_{u}, 0\}Z_{v_3}\!-\!Z_{x_3}Z_{x_2}^T\!=\!0;\ Z_{v_4}-\left[0 \  Z_{v_2}[I_{n_{o}}\quad 0]^T\  0\right]-\nonumber \\
&[Z_{x_3}\tilde{A}\quad 0]=0;
 Z_{v_5}-Z_{v_4}Z_{v_3}^T=0;
Z_{v_6}-Z_{x_3}Z_{x_5}^T=0; Z_{v_7}-\nonumber \\
&Z_{v_1}Z_{v_6}^T=0;\   Z_{v_8}-Z_{x_1}\mathrm{diag}\{\tilde{B}_{u}, 0\}=0;\
Z_{v_{9}}-Z_{v_8}Z_{x_7}^T=0;\nonumber \\
&Z_{v_{10}}\!\!-\!Z_{x_9}Z_{x_6}^T\!=\!0; Z_{v_{11}}\!\!-\!Z_{x_{1}}\!\mathrm{diag}\{\tilde{A}, 0\}\!\!-\!\left[0\   Z_{v_9}[I_{n_{f}}\  0]^T\  0\right]\!-\nonumber \\
&\left[\![Z_{x_4}\hat{A}_{uo}-Z_{v_{10}}G\tilde{C}_1\ \ 0]\   0\!\right]\!=\!0;
Z_{v_{12}}\!\!-\!\!Z_{v_{11}}Z_{v_3}^T\!=\!0;Z_{v_{13}}\!\!-\!\!Z_{v_{9}}\!\!-\nonumber \\
&Z_{x_4}\hat{A}\!-\!Z_{v_{10}}G\tilde{C}\!=\!0; Z_{v_{14}}\!\!\!-\!\!\!Z_{v_{13}}Z_{v_6}^T\!=\!0;  Z_{v_{15}}\!\!-\!\!Z_{x_1}\!Z_{x_5}^T\!=\!0; Z_{v_{16}}\!\!-\nonumber \\
&Z_{x_4}Z_{x_{8}}^T\!=\!0; Z_{v_{17}}\!-\!Z_{x_9}Z_{x_4}^T\!=\!0;   Z_{v_{18}}\!-\!Z_{x_2}Z_{x_1}^T\!=\!0;  Z_{v_{19}}\!\!-\!\!Z_{x_{3}}\!\times\nonumber \\
&Z_{v_{3}}^T=0; Z_{v_{15}}\!+\!Z_{v_{16}}\!=\!0;  Z_{v_{17}}\!+\!Z_{v_{8}}\!=\!I_{\hat{n}}; Z_{x_2}\!-\!Z_{x_2}^T\!=\!0;  Z_{x_8}\!-\nonumber \\
&Z_{x_8}^T\!=\!0;  Z_{x_5}\!\!-\!\!Z_{x_9}^T\!=\!0;  [I_{n}\  0]Z_{x_1}[0\  I_{\hat{n}-n}]^T\!\!\!=\!0;   [0\  I_{\hat{n}-n}]Z_{x_1}\!\!=\!0; \nonumber \\
&[I_{n}\  0]Z_{x_2}[0\  I_{\hat{n}-n}]^T\!=\!0;  [0\  I_{\hat{n}-n}]Z_{x_2}\!=\!0;  Z_{x_3}[0\ I_{\hat{n}-n}]^T\!=\!0; \nonumber \\
&[0\quad I_{\hat{n}-n}]Z_{x_4}\!=\!0;[0\quad I_{\hat{n}-n_{y_m}}]Z_{x_6}\!\!=0; Z_{x_7}[0\quad I_{\hat{n}-n_u}]^T\!=\!0;\nonumber \\
&[0\quad I_{\hat{n}-n}]Z_{x_9}=0;\ \ Z_{x_1}\!-\!Z_{x_1}^T\!=\!0;\end{align} and a rank constraint
\begin{align}\label{classical-n_oank-conditions}
\mathrm{rank}(Z)\leq \hat{n}.
\end{align}
Condition \eqref{corollary1-condition} is satisfied with
\begin{align}
&P=\left[
                                              \begin{array}{cc}
                                               \quad [I_{n}\quad 0]Z_{x_2}[I_{n}\quad 0]^T & [I_{n}\quad 0]Z_{x_5}^T \\
                                               Z_{x_5}[I_{n}\quad 0]^T & Z_{x_8}\\
                                              \end{array}
                                            \right]>0
\end{align} and  \eqref{LMI-condition-classical1} is satisfied with \!\!$\Omega_1\!=\!\left[I_{n} \ 0 \right]\!Z_{v_{11}}\!\left[I_{n} \ 0 \right]^T$, $\Omega_2=\left[I_{n} \
                                                                                                                  0 \right]Z_{v_{4}}\left[I_{n} \
                                                                                                                  0 \right]^T$, $\Omega_3=\left[I_{n}\ 0\right](Z_{v_{4}}-Z_{v_{7}})\left[I_{n} \
                                                                                                                  0 \right]^T$, $\Omega_4\!=\!\sqrt{2}\alpha\left[\!I_{n}\   0\right]\left(Z_{x_{1}}\left[I_{n} \
                                                                                                                  0 \right]^T\tilde{B}_f\!+\!Z_{v_{9}}\left[
                                                                                                                0\
                                                                                                                  I_{n_f} \right]^T\right)+\sqrt{2}\beta \left[I_{n}\   0\right]\times \left(\!\!Z_{x_4}\hat{B}_h\!+\!Z_{x_1}\!\!\left[\!
                                                                                                                \begin{array}{cc}
                                                                                                                  \!\!\tilde{B}_w \ \tilde{B}_u\!\!\\
                                                                                                                  \!\!0 \ \ \ 0\!\!\\
                                                                                                                \end{array}
                                                                                                              \!\right]\!\!+\!\!\left[\!Z_{x_4}\!\hat{B}_{w}\!\!-\!\!Z_{v_{10}}\!\left[\!
                                                                                                                                                      \begin{array}{c}
                                                                                                                                                         \!\!I_{n_{y_m}} \!\! \\
                                                                                                                                                      \!\! 0\!\! \\
                                                                                                                                                      \end{array}
                                                                                                                                                    \!\!\right]\!\!G\tilde{D}\  \ Z_{x_4}\!\tilde{B}_u\!\right]\!\right)$, $\Omega_5\!=\!Z_{x_3}\!\left[\!
                                                                                                                \begin{array}{cc}
                                                                                                                  \tilde{B}_w \ \tilde{B}_u\\
                                                                                                                  0 \ \ \ 0\\
                                                                                                                \end{array}
                                                                                                              \!\right]+\sqrt{2}\alpha \left(Z_{x_3}\left[I_n
                                                                                                                  \
                                                                                                                  0
                                                                                                              \right]^T\tilde{B}_f+Z_{v_2}\left[\!
                                                                                                                                            \begin{array}{c}
                                                                                                                                              \!\!0\!\!\\
                                                                                                                                              \!\!I_{n_f}\!\! \\
                                                                                                                                            \end{array}
                                                                                                                                          \!\right]\right)$, $\mathbf{A}=[I_{n}\quad 0](Z_{v_{12}}+Z_{v_{14}})$. We expect to find an estimation
error bound $\gamma$  close to $0$ to satisfy the following condition
\begin{align}\label{bound-2}
0<\mathrm{Tr}\left(Z_{x_8}\right)\leq\gamma.
\end{align}

If we can employ semi-definite programming to solve the
feasibility problem with the above constraints in which decision variables are the elements of $\mathbf{V}$,
then we have
\begin{align*}
L=Z_{x_{6}}^T\left[
                                                                                                                                      I_{n_{y_m}} \
                                                                                                                                      0
                                                                                                                                  \right]^T,\ K=[I_{n_u}\  0]Z_{x_{7}}^T.
\end{align*}

\begin{remark}
In numerical implementation, we first compare $n$ and $\hat{n}$ to choose \textbf{Case 1} or \textbf{Case 2}, and then set an estimation error upper bound $\gamma$. Now, we employ semi-definite programming to solve the feasibility problem with this $\gamma$. According to the result, we can adjust $\gamma$ to run the program again until an acceptable solution is obtained. Furthermore, we could develop effective algorithms to optimize $\gamma$, which is beyond the scope of this paper.
\end{remark}

\subsection{An illustrative example}\label{Sec:classical-example}
Consider a quantum optical plant with faults as follows
  \begin{align}\label{example-plant}
dx(t)=&\left[\begin{array}{cc}
                          -1& 0 \\
                          3  &-1\\
                        \end{array}
                      \right]x(t)dt+\left[\begin{array}{cc}0& 0\\
2& -1\end{array}\right]dw(t)+\nonumber\\
&\left[\begin{array}{cc} 2 &1\\4& 3
\end{array}\right]dy_u(t)+\left[
                                 \begin{array}{c}
                                   0 \\
                                   1 \\
                                 \end{array}
                               \right]
f(t)dt,\nonumber
\\dy(t)=&\left[
            \begin{array}{cc}
              -3 & 1 \\
              4 & -2 \\
            \end{array}
          \right]
x(t)dt+\left[
            \begin{array}{cc}
              1 & 0 \\
              0 & 1 \\
            \end{array}
          \right]dw(t),
 \end{align}
where the quantum plant matrices satisfy physical realizability conditions (i)-(iii) and $f(t)$ is represented as
\begin{equation}\nonumber
    f(t)=\begin{cases}
    0.25 \mathrm{cos}t,\quad \quad \quad\quad\quad\  0\leq t \leq 10,\\
    0.5+0.4 \mathrm{sin}(t-10),\quad 10<t \leq 20.
    \end{cases}
    \end{equation}

Applying the proposed  numerical procedure to the
quantum plant \eqref{example-plant}  for a given estimation error upper bound $\gamma=0.001$, we obtain
\begin{align*}
&L=\left[
                  \begin{array}{cc}
  -4.07   & 1.03\\
    9.22  & -30.21\\
                  \end{array}
                \right],
K=\left[
                  \begin{array}{cc}
    -0.1500  & -2.6643\\
   0.3000 &   2.6857\\
                  \end{array}
                \right].
\end{align*}
Now we check  that if the resulting solutions  satisfy the constraints listed
in \textbf{Problem} \textbf{1}.  It is easy to check that $P$ is a positive symmetric matrix and conditions \eqref{theorem1-error} and \eqref{corollary1-condition} are satisfied. Furthermore, $\mathrm{Tr}(Y_2)=0.001$. That means that we design an acceptable fault-tolerant controller for the system.

\section{Conclusion}\label{sec:conclusion}
In this note, we have investigated a quantum measurement-based feedback control system subject to fault signals. An estimator-based fault-tolerant controller has been designed to guarantee that the feedback control system with faults is stable.   Numerical procedures  have been  proposed for  fault-tolerant controller design. An example
was presented to test the proposed procedures. These results can provide
helpful guidelines for quantum optics experiments where  faults may occur in the quantum control systems.


\end{document}